# DDoS Attack and Defense Scheme in Wireless Ad hoc Networks

S.A.Arunmozhi[1], Y.Venkataramani[2]

[1.] Associate Professor,Dept. of ECE, Saranathan College of Engineering,India
[2].Principal, Saranathan College of Engineering, India

***Abstract.*** *The wireless ad hoc networks are highly vulnerable to distributed denial of service(DDoS) attacks because of its unique characteristics such as open network architecture, shared wireless medium and stringent resource constraints. These attacks throttle the tcp throughput heavily and reduce the quality of service(QoS) to end systems gradually rather than refusing the clients from the services completely. In this paper, we discussed the DDoS attacks and proposed a defense scheme to improve the performance of the ad hoc networks. Our proposed defense mechanism uses the medium access control (MAC) layer information to detect the attackers. The status values from MAC layer that can be used for detection are Frequency of receiving RTS/CTS packets, Frequency of sensing a busy channel and the number of RTS/DATA retransmissions. Once the attackers are identified, all the packets from those nodes will be blocked. The network resources are made available to the legitimate users. We perform the simulation with Network Simulator NS2 and we proved that our proposed system improves the network performance.*

## 1 Introduction

Wireless networks are inherently susceptible to security problems. The intrusion on the transmission medium is easier than for wired networks and it is possible to conduct denial of service attacks by scrambling the used frequency bands. The ad hoc context increases the number of potential security vulnerabilities. Ad hoc networks can not benefit from the security services offered by dedicated equipment such as firewalls, authentication servers and so on. The security services must be distributed, cooperative and consistent with the available bandwidth.

One of the serious attacks to be considered in ad hoc network is DDoS attack. A DDoS attack is a large-scale, coordinated attack on the availability of services at a victim system or network resource. The DDoS attack is launched by sending an extremely large volume of packets to a target machine through the simultaneous cooperation of a large number of hosts that are distributed throughout the network. The attack traffic consumes the bandwidth resources of the network or the computing resource at the target host, so that legitimate requests will be discarded.

A bandwidth depletion attack is designed to flood the victim network with unwanted traffic that prevents legitimate traffic from reaching the victim system. A resource depletion attack is an attack that is designed to tie up the resources of a victim system. This type of attack targets a server or process at the victim making it unable to legitimate requests for service. Any amount of resources can be exhausted with a sufficiently strong attack. The only viable approach is to design defense mechanism that will detect the attack and respond to it by dropping the excess traffic.

The DoS attacks that target resources can be grouped into three broad scenarios. The first attack scenario targets Storage and Processing Resources. This is an attack that mainly targets the memory, storage space, or CPU of the service provider. Consider the case where a node continuously sends an executable flooding packet to its neighborhoods and to overload the storage space and deplete the memory of that node. This prevents the node from sending or receiving packets from other legitimate nodes. The second attack scenario targets energy resources, specifically the battery power of the service provider. A malicious node may continuously send a bogus packet to a node with the intention of consuming the victim's battery energy and preventing other nodes from communicating with the node. The use of localized monitoring can help in detecting such nodes and preventing their consequences. The third attack scenario targets bandwidth. Consider the case where an attacker located between multiple communicating nodes wants to waste the network bandwidth and disrupt connectivity. This consumes the resources of all neighbors that communicate, overloads the network, and results in performance degradations. Such attacks can be prevented based on our proposed congestion based defense scheme.

# 2 Related Work

Xiapu Luo et al [1] have presented the important problem of detecting pulsing denial of service (PDoS) attacks which send a sequence of attack pulses to reduce TCP throughput. Wei-Shen Lai et al [3] have proposed a scheme to monitor the traffic pattern in order to alleviate distributed denial of service attacks. Shabana Mehfuz1 et al [4] have proposed a new secure power-aware ant routing algorithm (SPA-ARA) for mobile ad hoc networks that is inspired from ant colony optimization (ACO) algorithms such as swarm intelligent technique. Xiaoxin Wu et al [6] proposed a DoS mitigation technique that uses digital signatures to verify legitimate packets, and drop packets that do not pass the verification Ping Yi, Zhoulin Dai, Shiyong Zhang and Yiping Zhong [8] have presented a new DOS attack and its defense in ad hoc networks. The new DOS attack, called Ad Hoc Flooding Attack(AHFA), can result in denial of service when used against on-demand routing protocols for mobile ad hoc networks, such as AODV & DSR. John Haggerty, Qi Shi and Madjid Merabti [9] have proposed a new approach that utilizes statistical signatures at the router to provide early detection of flooding denial-of-service attacks. Wei Ren, Dit-Yan Yeung, Hai Jin, Mei Yang [11] have proposed a defense scheme that includes both the detection and response mechanisms. In this paper the detection scheme that monitors MAC layer signals and a response scheme based on Explicit Congestion Notification (ECN) marking are discussed. But, the method of monitoring the sending rates of the nodes is not discussed. Hence identifying the attacking nodes becomes a problem. It may also result in increase of false positives and false negatives. Gahng-Seop Ahn et al [12] have proposed SWAN, a stateless network model which uses distributed control algorithms to deliver service differentiation in mobile wireless ad hoc networks in a simple, scalable and robust manner. Giriraj Chauhan and Sukumar Nandi [13] proposed a QoS aware on demand routing protocol that uses signal stability as the routing criteria along with other QoS metrics.

# 3 Proposed Defense Technique

In this paper, we propose a new defense mechanism which consists of a flow monitoring table (FMT) at each node. It contains flow id, source id, packet sending rate and destination id. Sending rates are estimated for each flow in the intermediate nodes. The updated FMT is sent to the destination along with each flow. After monitoring the MAC layer signals the destination sends the Explicit Congestion Notification (ECN) bit to notify the sender nodes about the congestion. The sender nodes, upon seeing these packets with ECN marking, will then reduce their sending rate. If the channel continues to be congested because some sender nodes do not reduce their sending rate, it can be found by the destination using the updated FMT. It checks the previous sending rate of a flow with its current sending rate. When both the rates are same, the corresponding sender of the flow is considered as an attacker. Once the DDoS attackers are identified, all the packets from those nodes will be discarded.

The two phases of the proposed scheme are Bandwidth querying phase and Data transmission phase. In bandwidth querying phase the control messages sent are Bandwidth query request and Bandwidth query reply. The request packet includes the source IP address, destination IP address, type of the message, flow ID, and requested data rate which is stored in the bottleneck bandwidth (BnBW) field. In Bandwidth querying phase of the proposed scheme, the node's FMT information along a path is computed. An intermediate node updates its FMT using the BnBW value stored in the reply packet after receiving a REPLY message on the reverse path, and then forwards the REPLY to the next node. The available bandwidth *ABWj* is checked first. The reservation of bandwidth for the flow can be made if the value of *ABWj* is greater than or equal to the BnBW value in the REPLY packet. Else, the BnBW value in the REPLY packet is overwritten with the smaller value of *ABWj*. Next, the current BnBW value in the REPLY packet is added to the reserved rate *RRij*, associated with the in-out stream. A FMT entry is created with an assigned rate value A*Rij*, set equal to the BnBW value of the REPLY packet, if the stream (*i, j*) was previously inactive. Subsequently, the REPLY packet is forwarded to the next node on the reverse path. Finally, based on the value of the BnBW field, the source establishes the real-time flow when the REPLY packet reaches the source node.

### Distributed Rate Control

The actual rate $ACR_{ij}$ is given by

$$ACR_{ij}=W*AR_{ij} \tag{1}$$

Where,

$$W = \left[ \frac{Lc}{\sum ARij} \right]$$

where $L_c$ is the link capacity of each link.

If the source receives the congestion bit (CB), then the assigned rate $AR_{ij}$ can be written as

$$AR_{ij}=ACR_{ij}-\delta \tag{2}$$

Where δ is the rate reduction factor.

The rate monitoring function measures the traffic rate of a given in-out stream over a time interval T.

$$MR_{ij}=C_{ij}/T \tag{3}$$

If the measured rate $MR_{ij}$ is greater than actual rate $ACR_{ij}$ for the next time interval then the flow is considered to be an attack flow. Then its status is marked as REJECTED and the corresponding source IP address is recorded from the FMT.

## 4 Experimental Results

The network simulator NS2 is used to simulate our proposed algorithm. In our simulation, the channel capacity of mobile hosts is set to the same value: 2 Mbps. The distributed coordination function (DCF) of IEEE 802.11 is used for wireless LANs as the MAC layer protocol. It has the functionality to notify the network layer about link breakage. The simulated traffic is Constant Bit Rate (CBR).

Our simulation settings and parameters are summarized in table 1.

Table1: Simulation Parameters

| | |
|---|---|
| No. of Nodes | 80 |
| Area Size | 1200 X 1200 |
| Mac | 802.11 |
| Radio Range | 250m |
| Simulation Time | 60 sec |
| Traffic Source | CBR |
| Packet Size | 512 |
| Routing Protocol | AODV |

The experiment is carried out with five different normal flow of traffic with the data rate of 50 kbps and an attacking flow with the data rate of 500 kbps. The received bandwidth at the destination node over a period of time is obtained using the simulation and the results are shown in Fig. 1. The proposed system is compared with SWAN [2] and it is observed that the bandwidth received for the proposed system is greater compared to the other scheme. The bandwidth reservation technique and the distributed rate control technique of our proposed method make to achieve more bandwidth received for the legitimate users. Hence we are able to obtain greater bandwidth received for our proposed scheme than the SWAN.

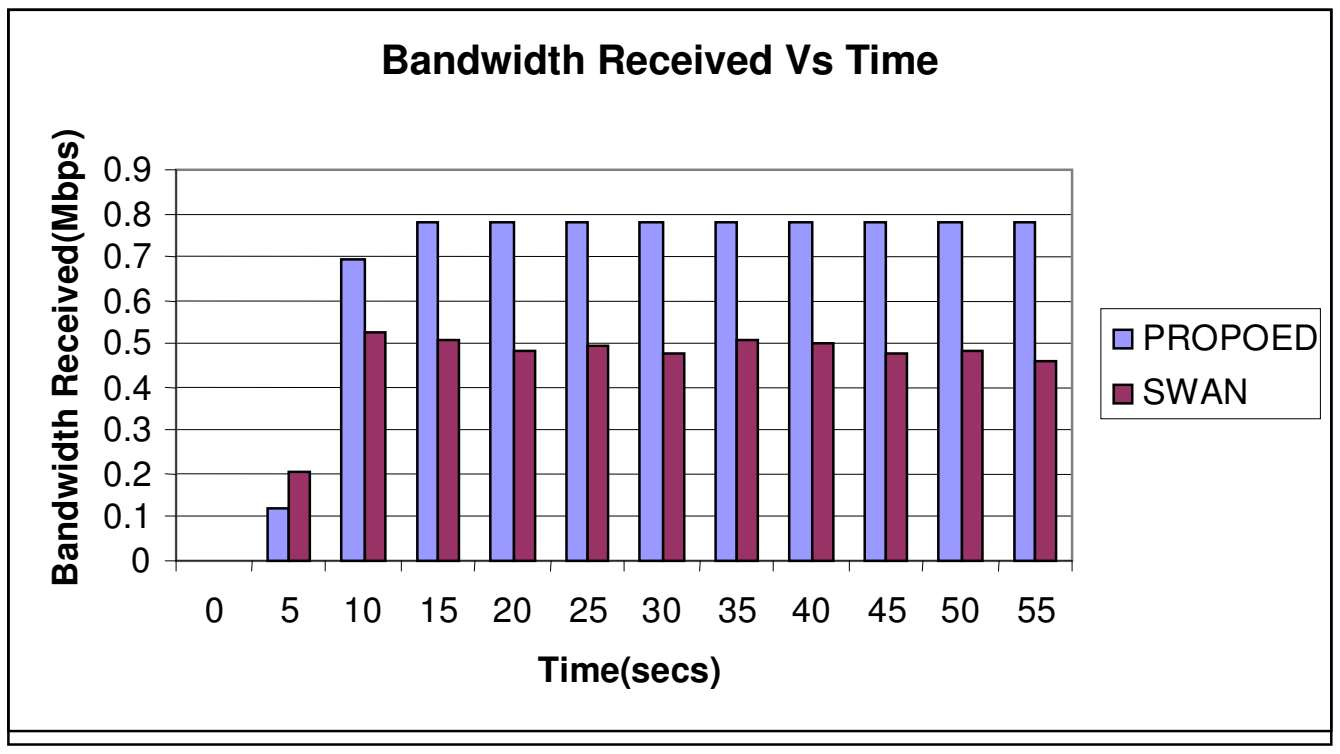

Fig. 1 Bandwidth Received Vs Time

The number of lost packets at the destination node over the period of time is shown in Fig. 2. From the graph, it is known that the number of packets lost for the proposed system is less compared to the SWAN. Since the attacking flows are rejected effectively in our proposed scheme, the network resources are made available to the legitimate users. This makes the dropping rate of legitimate packets low compared to the other scheme.

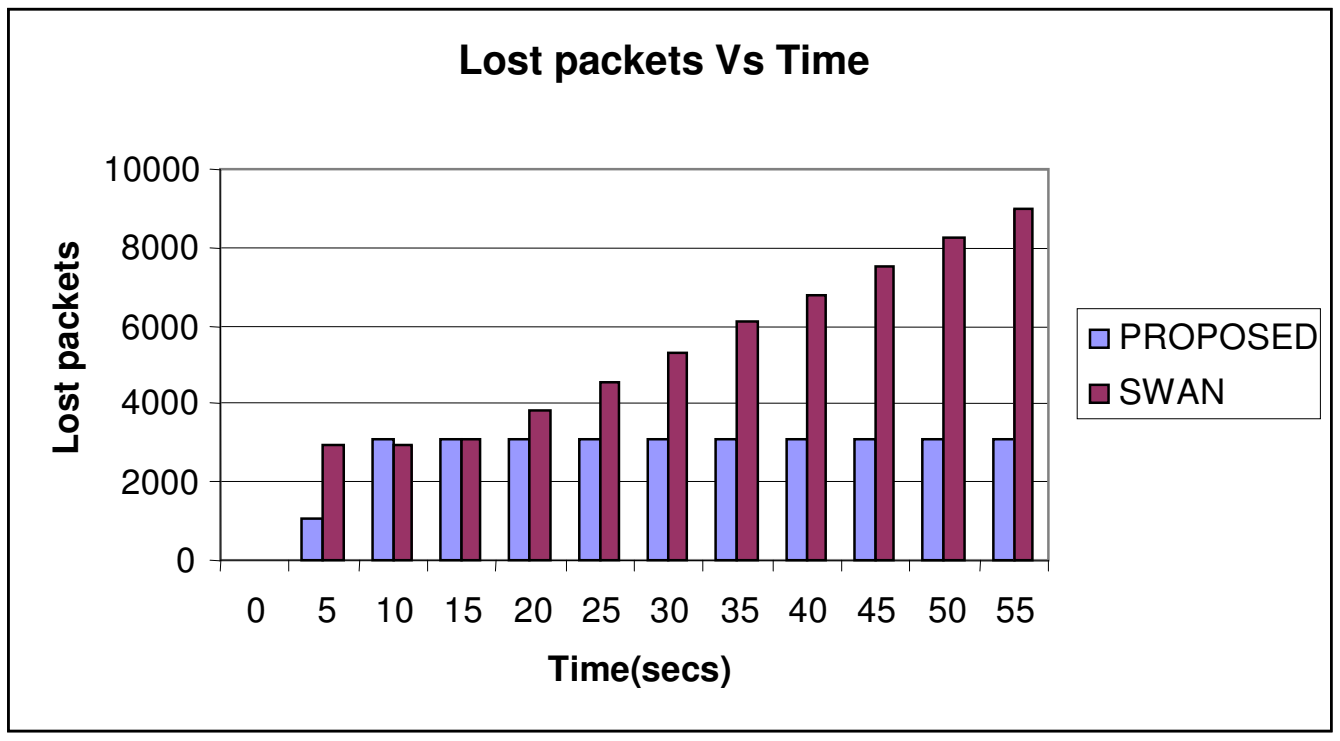

Fig. 2 Lost Packets Vs Time

The simulation is also carried with the variable number of attacking flows. Since the attacker are effectively identified and blocked in our scheme, it is again seen that the proposed scheme achieves greater received bandwidth. It is also observed that as the number of attackers is increased, the bandwidth received gets increased due to greater amount of traffic. The simulation results are shown in Fig. 3.

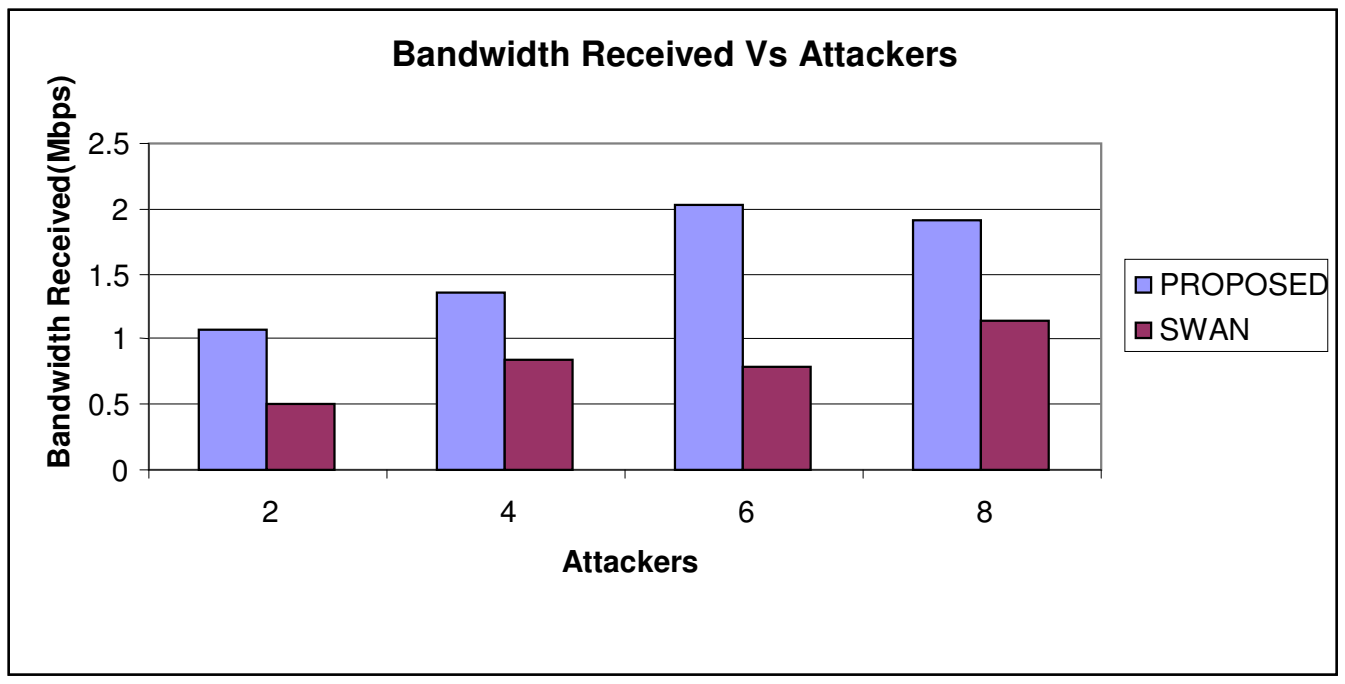

Fig. 3 Bandwidth Received Vs Attackers

The packet delivery ratio is the ratio of the number of packets received successfully and the total number of packets sent. Since the proposed scheme is congestion based scheme, the distributed rate control is applied when there is a heavy traffic due to the attacking flows. Hence the number of dropped packets is well reduced. As the dropped packets are reduced, more number of packets is delivered to the destination. Hence greater packet delivery ratio is achieved for the proposed scheme. The simulation results are shown in Fig. 4.

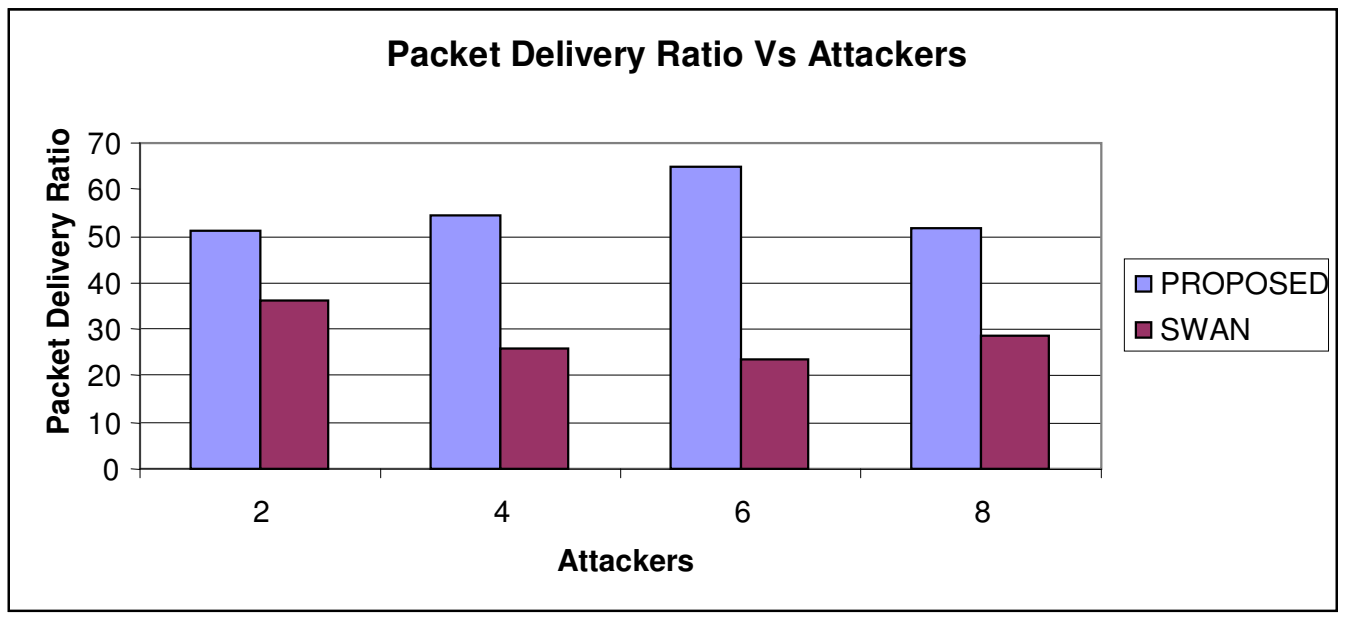

Fig. 4 Packet Delivery Ratio Vs Attackers

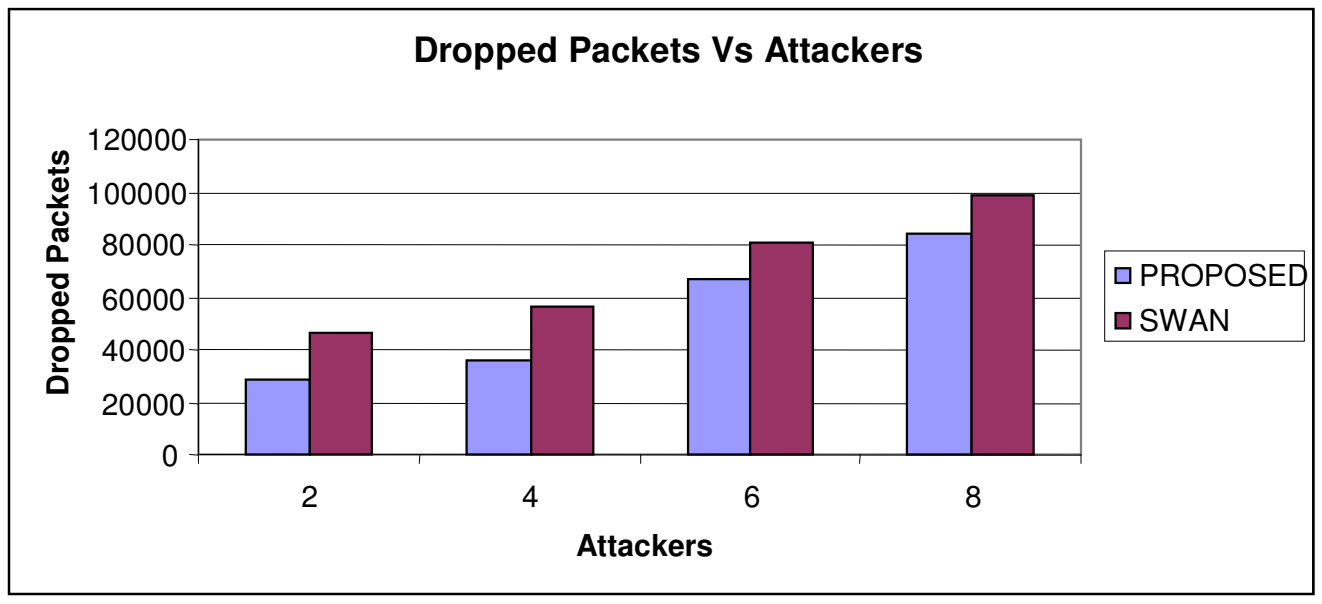

Fig. 5 Dropped Packets Vs Attackers

The number of dropped packets at the destination node with varying number of attackers is shown in Fig. 5. From the graph, it is known that the number of dropped packets for the proposed system is less compared to the SWAN. Since the attacking flows are rejected effectively, the dropping rate of legitimate packets low compared to the other scheme.

# 5 Conclusion

In this paper, we discussed the DDoS attacks and proposed a defense scheme to mitigate the attack in wireless ad hoc networks. Our approach can accurately identify DDoS attack flows and consequently apply rate-limiting to the malicious network flows. Our proposed defense mechanism identifies the attackers effectively. Once the attackers are identified, the attack traffic is discarded. This makes the network resources available to the legitimate users. We compared the performance of our proposed scheme with the SWAN scheme and proved that our proposed scheme assures better performance. By simulation results, we have shown that our proposed scheme achieves higher bandwidth received and packet delivery ratio with reduced packet drop for legitimate users.